\documentclass[12pt,leqno]{article}

\usepackage{amsmath,amssymb,amsfonts,theorem,latexsym}

\newtheorem{theo}{Theorem}[section]\newtheorem{lem}[theo]{Lemma}

\newtheorem{prop}[theo]{Proposition}

{\theorembodyfont{\rmfamily} \newtheorem{rem}[theo]{Remark}}
{\theorembodyfont{\rmfamily} }
{\theorembodyfont{\rmfamily} }
{\theorembodyfont{\rmfamily} }

\usepackage[all]{xy}

\DeclareFontFamily{U}{rsf}{}
\DeclareFontShape{U}{rsf}{m}{n}{
  <5> <6> rsfs5 <7> <8> <9> rsfs7 <10->  rsfs10}{}
\DeclareMathAlphabet{\mathscr}{U}{rsf}{m}{n}

\newcommand{\op}[1]{\operatorname{#1}}
\newcommand{\bGamma}{\boldsymbol{\Gamma}}

\newcommand{\bM}{\boldsymbol{M}}
\newcommand{\bpi}{\boldsymbol{\pi}}

\newcommand{\bL}{\boldsymbol{L}}

\newcommand{\ev}{\text{{\bfseries\sf ev}}}
\newcommand{\bT}{\boldsymbol{U}}
\newcommand{\bh}{\boldsymbol{h}}
\newcommand{\bSigma}{\boldsymbol{\Sigma}}
\newcommand{\bLambda}{\boldsymbol{\Lambda}}
\newcommand{\po}{p^{o}}
\newcommand{\bu}{\boldsymbol{u}}
\newcommand{\bq}{\boldsymbol{q}}
\newcommand{\bl}{\boldsymbol{l}}

\begin{document}

\title{Intermediate Jacobians and ${\sf {A}{D}{E}}$ Hitchin Systems}
\author{D.-E. Diaconescu$^{\flat}$, R. Donagi$^{\dagger}$, and 
T. Pantev$^{\dagger}$\\
$^\flat$ {\small New High Energy Theory Center, Rutgers University}\\
{\small 126 Frelinghuysen Road, Piscataway, NJ 08854}\\
$^{\dagger}$ {\small Department of Mathematics, University of Pennsylvania} \\
{\small David Rittenhouse Lab., 209 South 33rd Street,
Philadelphia, PA  19104-6395}\\
}
\date{July 2006}

\maketitle 

\begin{abstract} 
Let $\Sigma$ be a smooth projective complex curve and $\mathfrak{g}$ a 
simple Lie algebra of type ${\sf ADE}$ with associated adjoint group $G$. 
For a fixed pair $(\Sigma, \mathfrak{g})$, we  construct a family of 
quasi-projective Calabi-Yau  threefolds parameterized by the base of the 
Hitchin integrable system 
associated to $(\Sigma,\mathfrak{g})$. Our main result establishes an 
isomorphism between the Calabi-Yau integrable system, whose fibers are the
intermediate Jacobians of this family of Calabi-Yau threefolds, and
the Hitchin system for $G$, whose fibers are Prym varieties of the corresponding 
spectral covers. This 
construction provides a geometric framework for Dijkgraaf-Vafa transitions 
of type ${\sf ADE}$. In particular, it predicts an interesting connection 
between adjoint ${\sf ADE}$ Hitchin systems and quantization of holomorphic branes
on Calabi-Yau manifolds. 
\end{abstract}

\section{Introduction}

Large $N$ duality has been a central element in many recent
developments in topological string theory. {\bf A}-model large $N$
duality has led to exact results in the Gromov-Witten theory of
quasi-projective Calabi-Yau threefolds equipped with a torus action
\cite{amv,dfg1,dfg2,akmv,llz,lllz,dfs}. {\bf B}-model large $N$ duality
\cite{dv:matrix,dv:geom} predicts a very interesting relation between
matrix models, algebraic geometry and integrable systems, whose
mathematical structure has not been understood so far.  A first step
in this direction has been taken in \cite{dddhp}, where it was
recognized that {\bf B}-model large $N$ duality is intimately
connected to Hitchin integrable systems on projective curves.  More
precisely, the results of \cite{dddhp} relate the ${\sf A}_1$ Hitching
system on a smooth projective curve $\Sigma$ of arbitrary genus to the
large $N$ limit of a holomorphic brane system. A key element in
\cite{dddhp} is the construction of a family ${\mathcal X} \to {\bf
L}$ of quasi-projective Calabi-Yau threefolds so that the associated
intermediate Jacobian fibration $J^3({\mathcal X}/{\bf L})$ is
isogenous to the Prym fibration of the ${\sf A}_1$ Hitchin system on
$\Sigma$. Here ${\bf L}\simeq H^0(\Sigma, K_\Sigma^{\otimes 2})$ is
the base of the ${\sf A}_1$ Hitchin system.

The purpose of the present paper is to generalize this construction to
{\sf ADE} Hitchin systems on $\Sigma$. For a fixed simple Lie algebra
$\mathfrak{g}$ of type {\sf ADE}, we construct in
section~\ref{s:moduli} a family ${\mathcal X} \to {\bf L}$ of
quasi-projective Calabi-Yau threefolds parameterized by the base ${\bf
L}$ of the corresponding {\sf ADE} Hitchin system.  In
section~\ref{s:theorem} (Lemma~\ref{lem-pure}) we show that the
cohomology intermediate Jacobian $J^3(X)$ of a smooth generic
threefold in the family ${\mathcal X} \to {\bf L}$ is an Abelian
variety. Using the results of \cite{dm}, it follows that the associated
intermediate Jacobian fibration $J^3({\mathcal X}/{\bf L})$ is an
algebraic integrable system. For future reference, we will refer
to such integrable systems as Calabi-Yau integrable systems.

Our main result (Theorem~\ref{thm-main}, section~\ref{s:theorem}) states that
the intermediate Jacobian fibration $J^3({\mathcal X}/{\bf L})$ is
isomorphic to the Prym fibration of the {\sf ADE} Hitchin system for
the adjoint group $G$ associated to the Lie algebra
$\mathfrak{g}$. Moreover, this is an isomorphism of integrable
systems.

By analogy with the case of ${\sf A}_1$ Hitchin systems considered in
\cite{dddhp}, this result suggests that an arbitrary {\sf ADE} Hitchin
system is related to the large $N$ limit of a holomorphic brane
system. This question will be investigated elsewhere.

There are several possible generalizations of our results inspired by 
situations often encountered in physics. As explained in 
\cite[Section 3.1]{dddhp}, the geometric set-up considered in this paper 
is related by linearization to the deformation theory of projective 
Calabi-Yau threefolds with curves of singularities. 
The case considered in this paper corresponds to Calabi-Yau 
threefolds with a curve 
$\Sigma$ of split ${\sf {A}{D}{E}}$ singularities. There are two natural 
generalizations of this set-up. Namely, one can consider curves 
of ${\sf {A}{D}{E}}$ singularities with nontrivial monodromy
and one can also allow the singularity type to jump at special 
points on $\Sigma$. In the first case, we expect a relation between 
Calabi-Yau integrable systems and Hitchin systems of type 
${\sf {B} {C}{F}{G}}$. 
In the second case, we expect a similar relation between Calabi-Yau integrable 
systems and meromorphic Hitchin systems on $\Sigma$. Both situations  
are of physical interest and will be studied in future work. 
\medskip

\noindent {\it Acknowledgments.} We are very grateful to Robbert Dijkgraaf 
for many insightful discussions and collaboration in the related work 
\cite{dddhp}. We would also like to thank Bal\'{a}zs Szendr\"{o}i for help 
and encouragement. D.-E. D. has been partially supported by an Alfred P. Sloan 
fellowship and NSF grant PHY-0555374-2006. The work of R.D. 
was supported by NSF grants DMS 0104354 and DMS 0612992 and by NSF Focused Research Grant 
DMS 0139799, ``The Geometry of Superstrings''. The work of T.P. was supported by the NSF grant DMS 0403884 and NSF Focused Research Grant DMS 0139799.

\section{Moduli spaces} \label{s:moduli}

In this section we describe the relevant moduli spaces of non-compact
Calabi-Yau manifolds, as well as the corresponding universal families.
To set things up we fix the following data:
\begin{itemize}
\item a finite subgroup $\bGamma \subset \op{SL}(2,\mathbb{C})$,
corresponding via the McKay correspondence to a simple Lie algebra $\mathfrak{g}$ of
type {\sf ADE}. Write $\mathfrak{t} \subset \mathfrak{g}$ for a Cartan
subalgebra in $\mathfrak{g}$.
\item a fixed integer $g \geq 2$.
\item a smooth curve $\Sigma$ of genus $g$.
\item a $\bGamma$-equivariant rank two holomorphic vector bundle $V$
  on $\Sigma$
\item an isomorphism $\det V \cong K_{\Sigma}$.
\end{itemize}

\begin{rem} The existence of a $\bGamma$-equivariant structure can
  impose a constraint on $V$ \cite{balazs-fiber}:
\begin{description}
\item[Type $\text{\sf A}_{1}$:] If $\bGamma \leftrightarrow \text{\sf
  A}_{1}$, then $V$ is unconstrained.
\item[Type $\text{\sf A}_{n > 1}$:] If $\bGamma \leftrightarrow
\text{\sf A}_{n > 1}$, then we must have $V = L\oplus
(K_{\Sigma}\otimes L^{-1})$ for some line bundle $L$ on $\Sigma$. In
fact, $V$ will be polystable if and only if $\deg L = g-1$.
\item[Type {\sf D or E}:] If $\bGamma \leftrightarrow \text{\sf D}_{n
> 2}$ or $\bGamma \leftrightarrow \text{\sf E}_{6,7,8}$, then we must
have $V = \alpha \oplus \alpha$ for some line bundle $\alpha$ on
$\Sigma$ with $\alpha^{\otimes 2} = K_{\Sigma}$
\end{description}

\noindent
These restrictions follow immediately by noticing that
$\bGamma$-equivariance reduces the structure group of $V$ to the
centralizer of $\bGamma$ inside $\op{GL}(2,\mathbb{C})$. In the three
cases above this centralizer is $\op{GL}(2,\mathbb{C})$,
$\mathbb{C}^{\times}\times \mathbb{C}^{\times}$ and
$\mathbb{C}^{\times}$.
\end{rem}

\

\noindent
To this data we can associate the Calabi-Yau threefold
\[
X_{0} := \op{tot}(V)/\bGamma
\] 
which fibers over $\Sigma$ with fibers ALE spaces of type
$\bGamma$. This $X_{0}$ is the central fiber of a family $\mathcal{X}
\to \bL$ of non-compact Calabi-Yau threefolds that is constructed as
follows. Let $W$ denote the Weyl group of $\mathfrak{g}$ acting on the
Cartan algebra $\mathfrak{t}$. Consider the spaces
\[
\begin{split}
\widetilde{\bM} & = H^{0}(\Sigma, K_{\Sigma}\otimes \mathfrak{t}) \\
\bM & = H^{0}(\Sigma, K_{\Sigma}\otimes \mathfrak{t})/W \\ \bL & =
H^{0}(\Sigma, (K_{\Sigma}\otimes \mathfrak{t})/W).
\end{split}
\]

\begin{rem}
Notice that by construction $\bM$ and $\bL$ are affine algebraic
varieties equipped with a natural $\mathbb{C}^{\times}$-action coming
from the dilation action on the vector bundle $K_{\Sigma}\otimes
\mathfrak{t}$. In fact $\bL$ is isomorphic to a complex vector space
of dimension $(g-1)\dim(\mathfrak{g})$ with an appropriately defined
$\mathbb{C}^{\times}$-action. Indeed, the fiber bundle
$(K_{\Sigma}\otimes \mathfrak{t})/W$ can be identified with the
associated bundle $K_{\Sigma}^{\times}\times_{\mathbb{C}^{\times}}
\left( \mathfrak{t}/W \right)$. By Chevalley's theorem
\cite{chevalley} a choice of a basis of $W$-invariant homogeneous
polynomials on $\mathfrak{t}$ identifies the cone $\mathfrak{t}/W$
with a vector space of dimension $\op{rk}(\mathfrak{g})$ on which
$\mathbb{C}^{\times}$ acts with weights given by the degrees of the
polynomials in this basis. Thus $(K_{\Sigma}\otimes \mathfrak{t})/W$
can be viewed as a vector bundle of rank $\op{rk}(\mathfrak{g})$ on
$\Sigma$ and so $\bL$ is the vector space of global sections of this
vector bundle. Finally from the definition it is clear that the
natural map $\bM \to \bL$ is a closed immersion. This realizes $\bM$
as a closed subcone in $\bL$.
\end{rem}

\

\medskip

\noindent
Consider the $\op{rk}(\mathfrak{g}) + 1$ dimensional manifolds
\[
\begin{split}
\bT & := \xymatrix@1{\op{tot}((K_{\Sigma}\otimes \mathfrak{t})/W)
  \ar[r]^-{\bu} &   \Sigma},  \\
\widetilde{\bT} & := \xymatrix@1{\op{tot}(K_{\Sigma}\otimes
  \mathfrak{t}) \ar[r]^-{\tilde{\bu}} & 
\Sigma},
\end{split}
\]
and let $\bpi : \widetilde{\bT} \to \bT$ be the natural projection.
We will make frequent use of the following important proposition,
due to Bal\'{a}zs Szendr\"{o}i \cite{balazs-artin,balazs-fiber}:

\begin{prop} \begin{description}
\item[(a)] There exists a family of surfaces
$\bq : \mathcal{Q} \to \bT$, 
uniquely characterized by the properties
\begin{itemize}
\item $\mathcal{Q}_{\left| (\text{zero section of $\bT \to
      \Sigma$})\right.} \cong X_{0}$
\item $\mathcal{Q}_{|(\text{fiber of $\bT \to \Sigma$})} \cong \left(
      \text{the miniversal unfolding of $\mathbb{C}^{2}/\bGamma$} 
\right)$
\end{itemize}
\item[(b)] There exists a family $\hat{\bq} : \widehat{\mathcal{Q}} \to
  \widetilde{\bT}$ of smooth surfaces, together with a map
\[
\xymatrix@-1pc{ \widehat{\mathcal{Q}} \ar[rr]^{\varepsilon} \ar[rd] & &
\bpi^{*}\mathcal{Q} \ar[ld] \\ & \widetilde{\bT} & }
\]
which is a simultaneous resolution of all fibers of
$\bpi^{*}\mathcal{Q} \to \widetilde{\bT}$.
\item[(c)] For every section $\bl : \Sigma \to \bT$ of $\bu$, the
fiber product
\[
X_{\bl} := \mathcal{Q} \times_{\bq,\,\bT,\bl} \Sigma
\]
is a quasi-projective Gorenstein threefold with a trivial canonical
class.
\end{description}
\end{prop}
{\bf Proof.} This is proven in  \cite[Propositions~2.5, 2.7 and
  2.9]{balazs-artin} in a more general context by a cutting and
regluing argument. The idea is to build the family
$\widehat{\mathcal{Q}}$ from
copies of the Brieskorn-Grothendieck versal deformation of
$\mathbb{C}^{2}/\bGamma$ via the cocycle defining the
vector bundle $V$. In our simpler setup, one can also give a global
argument  by using the $\mathbb{C}^{\times}$ invariant Slodowy slice
\cite{slodowy} through a subregular nilpotent element. We will not
give the details here since this global construction is not essential
for our considerations.
\hfill $\Box$

\

\noindent
We can now construct our family $\mathcal{X} \to \bL$. Define
$\mathcal{X}$ as the pullback
\[
\mathcal{X} := \ev_{\bL}^{*}\mathcal{Q},
\]
where $\ev_{\bL} : \bL\times \Sigma \to \bT$ is the natural evaluation
map. The map $\mathcal{X} \to \bL$ is the composition $\mathcal{X} \to
\bL\times \Sigma \to \bL$.

Similarly we can construct a family $\widetilde{\mathcal{X}} \to
\widetilde{\bM}$, where $\widetilde{\bM} =
H^{0}(\Sigma,\widetilde{\bT})$ as the pullback
\[
\widetilde{\mathcal{X}} :=
\ev_{\widetilde{\bM}}^{*}\widehat{\mathcal{Q}},
\]
where $\ev_{\widetilde{\bM}} : \widetilde{\bM}\times \Sigma \to
\widetilde{\bT}$ is the natural evaluation map. The projection
$\widetilde{\mathcal{X}} \to \widetilde{\bM}$ is the composition
$\widetilde{\mathcal{X}} \to \widetilde{\bM}\times \Sigma \to
\widetilde{\bM}$.

\section{The main theorem} \label{s:theorem}

Note that the space $\bL$ which was defined as moduli of noncompact
Calabi-Yau manifolds has also an alternative description as moduli of
$\mathfrak{g}$ cameral covers of $\Sigma$ \cite{cameral}. Indeed, the 
reader will recognize $\bL = H^{0}(\Sigma,\bT)$ as the base of
the Hitchin system
\[
\bh : \op{Higgs}(\Sigma,G) \to \bL
\]
of topologically trivial $G$-Higgs bundles on $\Sigma$, where $G$ is
any complex Lie group with Lie algebra $\mathfrak{g}$. The case
relevant to us is when $G = G_{\op{ad}}$ is the adjoint form of
$\mathfrak{g}$. In fact we can construct the universal
$\mathfrak{g}$-cameral cover over $\bL$ as the pullback
\[
\widetilde{\bSigma} := \ev_{\bL}^{*}\widetilde{\bT}.
\]
The Hitchin fibration $\bh$ is known \cite{faltings,ron-dennis} to be
 a torsor over the relative Prym fibration
 $\op{Prym}_{G}(\widetilde{\bSigma}/\Sigma) \to \bL$ for the cameral
 covers.

Consider the discriminant locus $\Delta \subset \bL$. By this we mean
the locus of all $\ell \in \bL = H^{0}(\Sigma,\bT)$ which fail to be
transversal to the branch divisor of the cover $\widetilde{\bT} \to
\bT$. Outside of $\Delta$ both fibrations $\mathcal{X} \to \bL$ and
$\widetilde{\bSigma} \to \bL$ are smooth.  Fix $\ell \in \bL$ outside
of the discriminant. We get a smooth Calabi-Yau $\pi : X \to \Sigma$
and a smooth cameral cover $p : \widetilde{\Sigma} \to \Sigma$
corresponding to $\ell$. This geometry gives rise to two natural
complex abelian Lie groups: the cohomology intermediate Jacobian
$J^{3}(X)$ of $X$, and the Prym variety
$\op{Prym}_{G}(\widetilde{\Sigma},\Sigma)$ of the cover $p :
\widetilde{\Sigma} \to \Sigma$.

These groups have the following explicit description.  Let
$\bLambda_{G}$ be the group of cocharacters of the maximal torus
$\boldsymbol{T}_{G} \subset G$ of $G$. Since $G$ is adjoint,
$\bLambda_{G}$ is naturally the weight lattice of the Langlands 
dual Lie algebra ${^L}{\mathfrak{g}}$, which is identified with 
$\mathfrak{g}$ since our Lie algebra is simply laced. The
case of our theorem when $G = SO(3)$ was proven in \cite{dddhp}. From
now on we will therefore assume that $G$ is not of type ${\sf
A}_{1}$. In this case, the Prym is
\begin{equation} \label{eq:prym}
\op{Prym}_{G}(\widetilde{\Sigma},\Sigma) =
H^{1}\left(\Sigma,\left(p_{*} (\bLambda_{G}\otimes
\mathcal{O}_{\widetilde{\Sigma}}^{\times})\right)^{W}\right).
\end{equation}

More generally \cite{ron-dennis}, the Prym associated with a $G$
cameral cover, for any reductive $G$, is
$\op{Prym}(\widetilde{\Sigma},\Sigma) =
H^{1}(\Sigma,\mathcal{T}_{G})$, where $\mathcal{T}_{G}$ is a sheaf of
commutative groups on $\Sigma$ defined as
\[
\mathcal{T}_{G}(U) := \left\{ t \in \Gamma(p^{-1}(U),
\bLambda_{G}\otimes \mathcal{O}_{\widetilde{\Sigma}}^{\times})^{W}
\left|
\begin{minipage}[c]{1.5in} for every root $\alpha$ of $\mathfrak{g}$ we
  have $\alpha(t)_{|D^{\alpha}} = 1$ \end{minipage}\right. \right\}.
\]
In this formula we identify $\bLambda_{G}\otimes \mathbb{C}^{\times}$
with $\boldsymbol{T}_{G}$ and we view a root $\alpha$ as a
homomorphism $\alpha : \boldsymbol{T}_{G} \to
\mathbb{C}^{\times}$. The divisor $D^{\alpha} \subset
\widetilde{\Sigma}$ is the fixed divisor for the reflection
$s_{\alpha} \in W$ corresponding to $\alpha$. However, it was shown in
\cite[Theorem~6.5]{ron-dennis} that
\[
\mathcal{T}_{G} = \left(p_{*} (\bLambda_{G}\otimes
\mathcal{O}_{\widetilde{\Sigma}}^{\times})\right)^{W}
\]
as long as the coroots of the group $G$ are all primitive. As noted in
the introduction of \cite{ron-dennis}, this holds for all adjoint
groups of type {\sf ADE} {\em except} for our excluded case of
$SO(3)$. Therefore we have the identity \eqref{eq:prym}.

The intermediate Jacobians of $X$ are Hodge theoretic invariants of
 the complex structure of $X$. For general non-compact threefolds, 
they are generalized tori (= quotients of a vector space by a discrete 
abelian subgroup) defined in terms of
 the mixed Hodge structure on the cohomology or the homology of
 $X$. However in our case, the third cohomology of $X$ carries a 
pure Hodge structure as shown in the following lemma, so the 
intermediate Jacobians will be abelian varieties. 

\begin{lem} \label{lem-pure} Suppose that $X$ is a smooth non-compact
 Calabi-Yau threefold corresponding to $\ell \in \bL - \Delta$. Then
 the mixed Hodge structure on $H^{3}(X,{\mathbb Z})$ is pure of weight
 $3$ and of Hodge type $(1,2)+(2,1)$.
\end{lem}
{\bf Proof.}  To demonstrate the purity of the Hodge structure on
 $H^{3}(X,{\mathbb Z})$ we look at the map $\pi : X\to \Sigma$ onto
 the compact Riemann surface $\Sigma$. Let $\op{crit}(\pi) \subset
 \Sigma$ be the finite set of critical values of $\pi$. Set
 $\Sigma^{o} := \Sigma - \op{crit}(\pi)$, $X^{o} :=
 \pi^{-1}(\Sigma^{o})$, and let $j : \Sigma^{o} \hookrightarrow
 \Sigma$ and $\pi^{o} : X^{o} \to \Sigma^{o}$ denote the natural
 inclusion and projection maps. By the definition of $X$, explained in
 the previous section, the fibers of $\pi^{o}$ are complex surfaces
 isomorphic to smooth fibers of the universal unfolding of the
 singularity $\mathbb{C}^{2}/\Gamma$. In particular, the second
 homology of every fiber of $\pi^{o}$ is isomorphic to the root
 lattice of the Lie algebra $\mathfrak{g}$. By duality the second
 cohomology of every fiber of $\pi^{o}$ is isomorphic to the weight
 lattice of $\mathfrak{g}$. 
 Moreover, since all these fibers are deformation equivalent
 to the minimal resolution $\varepsilon :
 \widehat{\mathbb{C}^{2}/\bGamma} \to \mathbb{C}^{2}/\bGamma$ of
 $\mathbb{C}^{2}/\bGamma$, it follows that every fiber of $\pi^{o}$ is
 a deformation retract of the exceptional locus of $\varepsilon$ and
 so is homotopy equivalent to a configuration of $2$-spheres whose
 dual graph is the Dynkin diagram of $\mathfrak{g}$. This implies that
 for every $t \in \Sigma^{o}$ for the corresponding fiber $Q_{t} :=
 (\pi^{o})^{-1}(t)$ we have
\[
H^{0}(Q_{t},\mathbb{Z}) = \mathbb{Z}, \qquad H^{2}(Q_{t},\mathbb{Z}) =
\bLambda_{G},
\]
and the rest of the cohomology of $Q_{t}$ vanishes. 
A similar argument shows that the third cohomology group 
$H^3(Q_t,\mathbb{Z})$ also vanishes for singular fibers $Q_t$, 
with $t\in \op{crit}(\pi)$. Indeed, since the section $\ell \in H^0(\Sigma, 
{\bf U})$ is transversal to the branch divisor of the cover 
${\widetilde {\bf U}}\to {\bf U}$, the singular fibers of 
$\pi:X\to \Sigma$ have a single node. Therefore they are isomorphic to 
$\widehat{\mathbb{C}^2/\Gamma}$ with a $(-2)$ curve contracted. 
In particular the singular fibers have the homotopy type of a tree of rational 
curves, and their third cohomology vanishes. 

Therefore, by the Leray spectral sequence applied to the map 
$\pi : X \to \Sigma$ we get that
\[
H^{3}(X,\mathbb{C}) = H^{1}(\Sigma,R^{2}\pi_{*}\mathbb{C}) =
H^{1}(\Sigma,j_{*}R^{2}\pi^{o}_{*}\mathbb{C}).
\]
Next observe that the (a priori mixed) Hodge structure on the second cohomology
of each $Q_{t}$, $t \in \Sigma^{o}$ is pure and of type $(1,1)$, and
so $R^{2}\pi^{o}_{*}\mathbb{C}$ is a variation of pure Hodge
structures of Tate type and weight two. Indeed, it is obvious that the
Hodge structure on the second cohomology of
$\widehat{\mathbb{C}^{2}/\bGamma}$ is pure and of type $(1,1)$, since
the second homology of $\widehat{\mathbb{C}^{2}/\bGamma}$ is spanned by
the exceptional curves. The fact that $Q_{t}$ are all deformations of
the quasi-projective surface $\widehat{\mathbb{C}^{2}/\bGamma}$ and the
Gauss-Manin flatness of the weight filtration imply then that
$H^{2}(Q_{t},\mathbb{C})$ is pure and of type $(1,1)$. More
explicitly, by considering the versal deformation of the pair
consisting of the minimal resolution of $\mathbb{P}^{2}/\Gamma$ and
the divisor at infinity we can argue that each $Q_{t}$ is a rational
surface which admits a normal crossing compactification to a
projective rational surface $\overline{Q}_{t}$ with a tree $D_{t} =
\overline{Q}_{t} - Q_{t}$ of rational curves at infinity. Now writing
the relative cohomology sequence for $(\overline{Q}_{t},Q_{t})$ of the
pair and using the Gysin map we see that $H^{2}(Q_{t},\mathbb{C})$ is
of Tate type.

Finally, for any local system $\mathbb{L}$ of complex vector
spaces on $\Sigma^{o}$ 
with finite monodromy group $W$ we know that 
$H^{1}(\Sigma,j_{*}\mathbb{L})$ 
is the $W$-invariant subspace of
$H^{1}(\widehat{\Sigma},j_{*}\mathbb{\widehat{L}})$, 
where $\widehat{\Sigma}$ is the $W$-cover of $\Sigma$ determined 
by the monodromy, and ${\widehat{\mathbb{L}}}$ is the trivial local system
on $\widehat{\Sigma}^{o}$ which is the pullback of $\mathbb{L}$.
Now $H^{1}(\widehat{\Sigma},j_{*}\mathbb{\widehat{L}})$   
is the cohomology (with constant coefficients) of a smooth compact curve, so it
carries a pure Hodge structure of weight
$1$ and of Hodge type $(0,1)+(1,0)$, so the same applies to its 
$W$-invariant subspace $H^{1}(\Sigma,j_{*}\mathbb{L})$. In our case, this is 
(up to a Tate twist, i.e. shifting of all types by (1,1)) the same as
$H^{3}(X,\mathbb{C})$.
$\Box$

\

\bigskip

 Since $X$ is non-compact we will have to take extra care in
 distinguishing the intermediate Jacobians associated with the 
 Hodge structures on $H^{3}(X,{\mathbb Z})$ and $H_{3}(X,{\mathbb
 Z})$. We will denote these tori by $J^{3}(X)$ and
 $J_{3}(X)$ respectively. Explicitly
 \begin{align}
 J^{3}(X) & = H^{3}(X,{\mathbb C})/(F^{2}H^{3}(X,{\mathbb C}) +
   H^{3}(X,{\mathbb Z})), \label{eq:cohomology_jacobian}\\ J_{3}(X) &
   = H_{3}(X,{\mathbb C})/(F^{-1}H_{3}(X,{\mathbb C}) +
   H_{3}(X,{\mathbb Z})), \\ & = H^{3}(X,{\mathbb
   C})/(F^{2}H^{3}(X,{\mathbb C}) + H_{3}(X,{\mathbb Z})),
   \label{eq:homology_jacobian}
 \end{align}
where in the formula \eqref{eq:homology_jacobian} the inclusion
 $H_{3}(X,{\mathbb Z})/(\text{torsion}) \hookrightarrow
 H^{3}(X,{\mathbb C})$ is given by the intersection pairing map on
 three dimensional cycles in $X$.  More precisely, by the universal
 coefficients theorem we can identify $H^{3}(X,{\mathbb
 Z})/(\text{torsion})$ with the dual lattice $H_{3}(X,{\mathbb
 Z})^{\vee} := \op{Hom}_{{\mathbb Z}}(H_{3}(X,{\mathbb Z}),{\mathbb
 Z})$. Combining this identification with the intersection pairing on
 the third homology of $X$ we get a well defined map
 \[
 \xymatrix@R-2pc{i : \hspace{-2pc} & H_{3}(X,{\mathbb Z}) \ar[r] &
 H^{3}(X,{\mathbb Z})/(\text{torsion}) \\ & a \ar@{|->}[r] & \langle
 a,\bullet \rangle }
 \]
 which is injective on the free part of $H_{3}(X,{\mathbb
   Z})$. Combining $i$ with the natural inclusion $H^{3}(X,{\mathbb
   Z})/(\text{torsion}) \subset H^{3}(X,{\mathbb C})$ we obtain the
   map appearing in \eqref{eq:homology_jacobian}. Furthermore since
   $i$ is injective modulo torsion, it follows that the induced
   surjective map on intermediate Jacobians
 \begin{equation}
 J_{3}(X) \to J^{3}(X) \label{eq:compare_jacobians}
 \end{equation}
 is a finite isogeny of tori. Note that had $X$ been compact,
 the unimodularity of the Poincare pairing would have implied that
 \eqref{eq:compare_jacobians} is an isomorphism and so we would not have
 had to worry about the distinction between $J_{3}(X)$ and $J^{3}(X)$. 

By the  previous lemma it follows that if we twist the Hodge
structure on $H^{3}(X,\mathbb{C})$ by a Tate Hodge structure of weight
$(-2)$ we will get a pure effective Hodge structure of weight $1$. In
particular $J^{3}(X)$ and $J_{3}(X)$ are both abelian varieties which
are dual to each other. The lemma also implies that
 \[
 \begin{split}
 J_{3}(X) & = H_{3}(X,{\mathbb Z})\otimes_{{\mathbb Z}} S^{1} \\
 J^{3}(X) & = H^{3}(X,{\mathbb Z})\otimes_{{\mathbb Z}} S^{1}
 \end{split}
 \]
as real tori. Furthermore the isogeny \eqref{eq:compare_jacobians} can
 be identified explicitly as
 \[
 \xymatrix@R-1pc{ J_{3}(X) \ar[r] \ar@{=}[d] & J^{3}(X) \ar@{=}[d] \\
 H_{3}(X,{\mathbb Z})\otimes_{{\mathbb Z}} S^{1} \ar[r]_-{i\otimes
 \op{id}} & H^{3}(X,{\mathbb Z})\otimes_{{\mathbb Z}} S^{1} }
 \]
\

\bigskip

Our main result is
 
\begin{theo} \label{thm-main} Suppose $G$ is the adjoint Lie group with
  Lie algebra $\mathfrak{g}$. Away from the discriminant, the relative
  Prym fibration $\op{Prym}_{G}(\widetilde{\bSigma}/\Sigma) \to \bL$
  is isomorphic to the cohomology intermediate Jacobian fibration
  $J^{3}(\mathcal{X}/\bL) \to \bL$ for the family $\mathcal{X} \to
  \bL$. This isomorphism identifies the symplectic structure on
  Hitchin's space $\op{Prym}_{G}(\widetilde{\bSigma}/\Sigma)$ with the
  Poisson structure on $J^{3}(\mathcal{X}/\bL)$ coming from the Yukawa
  cubic on $\bL$ \cite{dm}.
\end{theo}
{\bf Proof.} First we show that the relative Prym fibration is
isomorphic to the cohomology intermediate Jacobian fibration as
families of polarized abelian varieties.

We divide the proof of this fact into three steps:

\

\medskip

\noindent
{\bf Step 1.} \ $J^{3}(X) = H^{3}(X,S^{1}) \cong
H^{1}(\Sigma,(p_{*}\bLambda_{G})^{W}\otimes S^{1})$. Indeed, if $\pi :
X \to \Sigma$ is the natural map, then we have $R^{1}\pi_{*}S^{1} = 1$
and $R^{3}\pi_{*}S^{1} = 1$. This follows from the explicit
description of the homotopy type of a smooth fiber $Q_{t}$ of $\pi$
given in the proof of Lemma~\ref{lem-pure}. By the Leray spectral
sequence we get $H^{3}(X,S^{1}) = H^{1}(\Sigma,
R^{2}\pi_{*}S^{1})$. Furthermore
\[
R^{2}\pi_{*}S^{1} \cong (R^{2}\pi_{*}\mathbb{Z})\otimes S^{1} \cong
(p_{*}\bLambda_{G})^{W}\otimes S^{1}.
\]
The first isomorphism follows from the universal coefficients
spectral sequence \cite{iversen} and the divisibility of $S^{1}$. The
identification $R^{2}\pi_{*}\mathbb{Z} \cong (p_{*}\bLambda_{G})^{W}$
over $\Sigma$ follows from the corresponding identification over $\bT$
which is classical.

\

\medskip

\noindent
{\bf Step 2.} \ $H^{1}(\Sigma, (p_{*}\bLambda_{G})^{W}\otimes S^{1})
\cong H^{1}(\Sigma, (p_{*}(\bLambda_{G}\otimes S^{1}))^{W})$.
 
In fact we will show:

\begin{lem}
If $p : \widetilde{\Sigma} \to \Sigma$ has simple Galois ramification,
then the natural map
\[
\nu : (p_{*}\bLambda_{G})^{W}\otimes S^{1} \to
(p_{*}(\bLambda_{G}\otimes S^{1}))^{W}
\]
is an isomorphism.
\end{lem}
{\bf Proof.}  Since $\nu$ is tautologically an isomorphism away from
the branch locus of $p : \widetilde{\Sigma} \to \Sigma$, we need only
check that $\nu$ is an isomorphism of stalks at the branch points of
$p$.

Suppose $b \in \Sigma-\Sigma^{o}$ is a branch point. We have
\begin{equation} \label{eq:stalks}
\begin{split}
((p_{*}\bLambda_{G})^{W})_{b} & \cong \bLambda_{G}^{\rho_{\alpha}} \\
((p_{*}(\bLambda_{G}\otimes S^{1}))^{W})_{b} & \cong
(\bLambda_{G}\otimes S^{1})^{\rho_{\alpha}}.
\end{split}
\end{equation}
where $\rho_{\alpha} : \bLambda_{G} \to \bLambda_{G}$ is the
reflection corresponding to a root
$\alpha$. Indeed observe  that
\[
\begin{split}
p_{*}\bLambda_{G} & = i_{*}(\po_{*}\bLambda_{G}) \\
p_{*}(\bLambda_{G}\otimes S^{1}) & = i_{*}(\po_{*}(\bLambda_{G}\otimes
S^{1})),
\end{split}
\]
where
\[
\xymatrix@C-2pc{ \widetilde{\Sigma} \ar[d]_-{p} & \supset &
\widetilde{\Sigma}^{o} \ar[d]^-{\po} \\ \Sigma & \supset & \Sigma^{o}
}
\]
and $\po$ denotes the part of $p$ away from ramification. In
particular if $x \in \Sigma^{o}$ is a point near $b$, we have that
\[
(p_{*}\bLambda_{G})_{b} = (p_{*}\bLambda_{G})_{x}^{\op{mon}_{x}} =
(\op{Fun}(p^{-1}(x),\bLambda_{G}))^{\op{mon}_{x}} =
(\op{Fun}(W,\bLambda_{G}))^{\op{mon}_{x}} =
\op{Fun}(W/s_{\alpha},\bLambda_{G}).
\]
Therefore
\[
(p_{*}\bLambda_{G})_{b}^{W} = \op{Fun}(W/s_{\alpha},\bLambda_{G})^{W}
= \bLambda_{G}^{\rho_{\alpha}}.
\]
An analogous argument gives the second identity in \eqref{eq:stalks}.

Thus our lemma is equivalent to showing that the natural map
\[
\nu_{b} : \bLambda_{G}^{\rho_{\alpha}}\otimes S^{1} \to
  (\bLambda_{G}\otimes S^{1})^{\rho_{\alpha}}
\]
is an isomorphism for one (hence all) roots $\alpha$.

We will analyze the {\sf ADE} types separately.

Suppose $G$ is of type ${\sf A}_{n}$. The short exact sequence
\[
1 \to GL(1) \to GL(n+1) \to \mathbb{P}GL(n+1) \to 1
\]
induces a short exact sequence of cocharacter lattices
\[
0 \to \bLambda_{GL(1)} \to \bLambda_{GL(n+1)} \to
\bLambda_{\mathbb{P}GL(n+1)} \to 0.
\]
Explicitly this is the sequence
\[
0 \to \mathbb{Z} \to \mathbb{Z}^{n+1} \to \bLambda_{\mathbb{P}GL(n+1)}
\to 0,
\]
where the map $\mathbb{Z} \to \mathbb{Z}^{n+1}$ is given by $1 \mapsto
(1,1,\ldots,1)$.

Choose $\alpha = [(0,0,\ldots,0,1,-1)]$. Then $\rho_{\alpha}$ is the
transposition
\[
(\lambda_{1},\ldots, \lambda_{n},\lambda_{n+1}) \to
(\lambda_{1},\ldots, \lambda_{n+1},\lambda_{n})
\] 
and so $[( \lambda_{1},\ldots, \lambda_{n},\lambda_{n+1})] \in
\bLambda_{\mathbb{P}GL(n+1)}$ is in
$\bLambda_{\mathbb{P}GL(n+1)}^{\rho_{\alpha}}$ if and only if
$\lambda_{n} = \lambda_{n+1}$. Similarly $\boldsymbol{x} := [(x_{1},
\ldots, x_{n+1})] \in \bLambda_{\mathbb{P}GL(n+1)}\otimes S^{1}$
satisfies $\rho_{\alpha}(\boldsymbol{x}) = \boldsymbol{x}$ if and only
if
\[
(1, \ldots,1,x_{n}x_{n+1}^{-1},x_{n}^{-1}x_{n+1}) = (x,\ldots,x,x,x)
\]
for some $x \in S^{1}$. Since by assumption $n > 1$ we see that
$\nu_{b}$ is surjective, hence an isomorphism. Note that in the
excluded case $n = 1$ the map $\nu_{b}$ has a non-trivial cokernel
$\mathbb{Z}/2$.

If $G$ is of type ${\sf D}_{n}$, $n > 2$, we use the basic sequence
\[
0 \to \bLambda_{SO(2n)} \to \bLambda_{\mathbb{P}SO(2n)} \to
\mathbb{Z}/2 \to 0
\]
and the fact that $\bLambda_{SO(2n)}$ can be identified with the
square lattice $\mathbb{Z}^{n}$. After tensoring with $S^{1}$ we get
\[
1 \to \{ \pm 1 \} \to \bLambda_{SO(2n)}\otimes S^{1} \to
\bLambda_{\mathbb{P}SO(2n)} \otimes S^{1} \to 1
\]
where the map $\{ \pm 1 \} \to \bLambda_{SO(2n)}\otimes S^{1} =
(S^{1})^{n}$ sends $(-1)$ to $(-1,-1, \ldots, -1)$. Since we are
assuming that $n > 2$, this yields the surjectivity of
$\nu_{b}$. Again in the excluded case of $n = 2$ we have
$\op{coker}(\nu_{b}) = \mathbb{Z}/2$.

Finally if $G$ is of type ${\sf E}_{n}$, for $n = 6,7,8$, then
$\bLambda_{G}$ can be identified with the quotient:
\[
0 \to \mathbb{Z} \to H^{2}(dP_{n},\mathbb{Z}) \to \bLambda_{G} \to 0,
\]
where $dP_{n}$ is a general del Pezzo surface of degree $9-n$ and
under the map $\mathbb{Z} \to H^{2}(dP_{n},\mathbb{Z})$, the generator
$1 \in \mathbb{Z}$ goes to the anti-canonical class $3\ell -
\sum_{i=1}^{n} e_{i}$. Here we think of $dP_{n}$ as the blow-up of
$\mathbb{P}^{2}$ at $n$ general points. Its cohomology has an
orthogonal basis $\{ \ell, e_{1}, \ldots, e_{n} \}$ satisfying
$\ell^{2} = 1$ and $e_{i}^{2} = -1$. Geometrically, $\ell$ is the
pullback of the hyperplane class on $\mathbb{P}^{2}$ and the $e_{i}$'s
are the exceptional curves.

For our root $\alpha$ we take $\alpha = e_{1} - e_{2}$. Then
$\rho_{\alpha}$ acts as the Picard-Lefschetz reflection
$\rho_{\alpha}(x) = x + \langle x, e_{1} - e_{2} \rangle \cdot (e_{1}
-e_{2})$. In particular $\rho_{\alpha}$ interchanges $e_{1}$ with
$e_{2}$ and fixes the rest of the basis. Now the same reasoning as
above shows that $\nu_{b}$ is surjective and hence an isomorphism.

\

\medskip

\noindent
{\bf Step 3.} The inclusion $S^{1} \subset \mathbb{C}^{\times}$ of
groups induces a natural inclusion of sheaves
\begin{equation} \label{eq-inclusion}
\iota : \bLambda_{G} \otimes S^{1} \hookrightarrow \bLambda_{G}\otimes
\mathcal{O}^{\times}_{\widetilde{\Sigma}}.
\end{equation} 
We claim that $\iota$ induces an isomorphism of tori
\[
h^{1}(\iota) : H^{1}(\Sigma, (p_{*}(\bLambda_{G}\otimes S^{1}))^{W})
\; \widetilde{\to} \; H^{1}(\Sigma, (p_{*}(\bLambda_{G}\otimes
\mathcal{O}^{\times}_{\widetilde{\Sigma}}))^{W}).
\]
Indeed, observe that $H^{1}(\Sigma, (p_{*}(\bLambda_{G}\otimes
S^{1}))^{W})$ is isogenous to
$H^{1}(\widetilde{\Sigma},\bLambda_{G}\otimes S^{1})^{W}$ and
similarly $H^{1}(\Sigma, (p_{*}(\bLambda_{G}\otimes
\mathcal{O}^{\times}_{\widetilde{\Sigma}}))^{W})$ is isogenous to
$H^{1}(\widetilde{\Sigma},\bLambda_{G}\otimes
\mathcal{O}_{\widetilde{\Sigma}}^{\times})^{W}$. Under these isogenies
the map $h^{1}(\iota)$ is compatible with the map
\[
H^{1}(\widetilde{\Sigma},\bLambda_{G}\otimes S^{1})^{W} \to
H^{1}(\widetilde{\Sigma},\bLambda_{G}\otimes
\mathcal{O}_{\widetilde{\Sigma}}^{\times})^{W}
\]
and so $h^{1}(\iota)$ is surjective with at most a finite kernel.

Let $C$ be the cone of the map of sheaves \eqref{eq-inclusion}.  Since
the constant sheaf $\mathbb{C}^{\times}_{\widetilde{\Sigma}}$ has a
resolution
\[
\mathbb{C}^{\times}_{\widetilde{\Sigma}} \to
\mathcal{O}^{\times}_{\widetilde{\Sigma}} \to
\Omega^{1}_{\widetilde{\Sigma}},
\]
and since $\mathbb{C}^{\times} = S^{1}\times \mathbb{R}$, it follows
that $C$ is quasi-isomorphic to a complex of $\mathbb{R}$-vector
spaces on $\widetilde{\Sigma}$ with cohomology sheaves
$\mathcal{H}^{0}C \cong \bLambda_{G}\otimes
\Omega^{1}_{\widetilde{\Sigma}}$ (considered as a sheaf of
$\mathbb{R}$-vector spaces, and $\mathcal{H}^{1}C \cong
\bLambda_{G}\otimes \mathbb{R}$. This implies that
\[
\op{cone}\left[ (p_{*}(\bLambda_{G}\otimes S^{1}))^{W} \to
(p_{*}(\bLambda_{G}\otimes
\mathcal{O}^{\times}_{\widetilde{\Sigma}}))^{W}\right]
\]
is a complex of sheaves of $\mathbb{R}$-vector spaces on $\Sigma$ and
so its hypercohomology can not be a torsion group. This implies that
$h^{1}(\iota)$ is injective and finishes the proof of the
identification of the cameral Pryms with the intermediate Jacobians.

\

\medskip

\noindent
To finish the proof of the theorem, it remains to show that the family
$\mathcal{X} \to \bL$ of non-compact Calabi-Yau manifolds gives rise
to a Yukawa cubic field on $\bL$, which coincides with the cubic
defining the symplectic structure \cite{dm} on the Higgs
moduli space. This is equivalent to showing that for a smooth
Calabi-Yau $X$ in $\bL$, there is a unique up to scale non-vanishing
holomorphic three form $\Omega$ on $X$, which is compatible with the
Seiberg-Witten $\mathfrak{t}$-valued one form $\eta$ on the corresponding
cameral cover $\widetilde{\Sigma}$. 

First we recall that the cameral cover $\widetilde{\Sigma}$ was
defined as the pullback of the cover $\widetilde{\bT} \to \bT$ via a a
map $\Sigma \to \bT$. In this picture, the Seiberg-Witten
$\mathfrak{t}$-valued holomorphic one form $\eta$ on $\widetilde{\Sigma}$
becomes simply the pullback of  the tautological section of the
pullback of $\mathfrak{t}\otimes K_{\Sigma}$ to $\widetilde{\bT} =
\op{tot}(\mathfrak{t}\otimes K_{\Sigma})$.

Next observe that as long as $V$ is semistable, the singular
Gorenstein Calabi-Yau $X_{0} = \op{tot}(V)/\Gamma$ has a unique up to
scale non-vanishing holomorphic three form $\Omega_{0}$. Indeed, the
ratio of any two such forms will be a global holomorphic function on
$X_{0}$, and so will pullback to a global holomorphic function on
$\op{tot}(V)$. But every such function can be written as a convergent
series of functions which are polynomial along the fibers of
$\op{tot}(V) \to \Sigma$. However these polynomial functions can be
thought of as sections in $S^{\bullet}V^{\vee}$, and since $V$ is
semistable of positive degree, it follows that all such sections are
constants. By semicontinuity this implies that for $\bl \in \bL$ in a
small neighborhood of zero, the Calabi-Yau manifold $X_{\bl}$ has a
unique up to scale holomorphic three form $\Omega_{\bl}$. Since our
universal family $\mathcal{X} \to \bL$ is preserved by the natural
action of $\mathbb{C}^{\times}$ on $\bL$, this shows that
$\Gamma(X_{\bl},\Omega^{3}_{X_{\bl}}) = \mathbb{C}$ for all $\bl$ in
$\bL$.

Let now $\bl \in \bL - \Delta$ and let $\pi : X \to \Sigma$ and $p :
\widetilde{\Sigma} \to \Sigma$ be the corresponding Calabi-Yau
threefold and cameral cover. We have a commutative diagram of spaces
\[
\xymatrix@R-1pc{
\widehat{X} \ar[rd]^-{\hat{\pi}} \ar[d]_-{\varepsilon} \\
X\times_{\Sigma}
\widetilde{\Sigma} \ar[r]^-{\tilde{\pi}} \ar[d]_-{f}
& \widetilde{\Sigma} \ar[d]^-{p} \\
X \ar[r]_-{\pi} & \Sigma
}
\]
where $\widehat{X} := \widehat{\mathcal{Q}}\times_{\widetilde{\bT}}
\widetilde{\Sigma}$ is the small resolution of $X\times_{\Sigma}
\widetilde{\Sigma}$, induced from $\widehat{\mathcal{Q}} \to
\mathcal{Q}$. Note that in this diagram $X\times_{\Sigma}
\widetilde{\Sigma}$ is Gorenstein and all
the other spaces are smooth.

Let $K_{\pi}$, $K_{\tilde{\pi}}$, $K_{\hat{\pi}}$ denote the relative
canonical classes of the morphisms $\pi$, $\tilde{\pi}$, $\hat{\pi}$.
Since the square in the above diagram is a fiber square we have
$K_{\tilde{\pi}} = f^{*}K_{\pi}$. Since $X\times_{\Sigma}
\widetilde{\Sigma}$ is Gorenstein and
$\varepsilon : \widehat{X} \to X\times_{\Sigma}
\widetilde{\Sigma}$ is small, we have that
$K_{\widehat{X}} = \varepsilon^{*}K_{X\times_{\Sigma}
\widetilde{\Sigma}}$ and therefore $K_{\hat{\pi}} =
\varepsilon^{*}K_{\tilde{\pi}}$. This gives an identification
\[
(f\circ \varepsilon)^{*}K_{X} = K_{\hat{\pi}}\otimes (p\circ
\hat{\pi})^{*} K_{\Sigma}.
\]
Let $\Omega$ denote the unique up to scale non-vanishing holomorphic
three form on $X$. Then $\Omega$ is a global nowhere vanishing section
of $K_{X}$ and so $(f\circ \varepsilon)^{*}\Omega$ is a non-vanishing
section of $K_{\hat{\pi}}\otimes (p\circ \hat{\pi})^{*} K_{\Sigma} =
\Omega^{2}_{\hat{\pi}}\otimes (p\circ \hat{\pi})^{*}
\Omega^{1}_{\Sigma}$. Using the section $\widehat{\Omega} := (f\circ
\varepsilon)^{*}\Omega$ we can construct a period map from
$\widetilde{\Sigma}$ to the total space of $\mathfrak{t}\otimes
p^{*}K_{\Sigma}$. Indeed, fix a base point $o \in \widetilde{\Sigma}$
and an identification $H^{2}(\widehat{X}_{o},\mathbb{C}) \cong
\mathfrak{t}$. Let $s \in \widetilde{\Sigma}$ and let $v \in
(p^{*}K_{\Sigma}^{-1})_{s}$.  The contraction of
$\widehat{\Omega}_{|\widehat{X}_{s}}$ with $\hat{\pi}^{*}(v)$ is a
closed two form on $\widehat{X}_{s}$ which can be transported by the
Gauss-Manin connection along a path from $s$ to $o$ to give an element
in $\mathfrak{t} = H^{2}(\widehat{X}_{o},\mathbb{C})$. Since by
construction $R^{2}\hat{\pi}_{*}\mathbb{C}$ is a trivial local system
on $\widetilde{\Sigma}$, this construction is independent of the
choice of a path and gives a well defined map $p^{*}K_{\Sigma}^{-1}
\to \mathfrak{t}\otimes \mathcal{O}_{\widetilde{\Sigma}}$ of
holomorphic vector bundles, or equivalently a holomorphic section
$\hat{\eta}$ in $\mathfrak{t}\otimes p^{*}K_{\Sigma}$ on
$\widetilde{\Sigma}$. Finally, to show that $\hat{\eta}$ coincides
with the Seiberg-Witten form, note that the period map $\hat{\eta}$ is
the composition of the inclusion $\widetilde{\Sigma} \hookrightarrow
\widetilde{U}$ and the universal period map $\hat{\underline{\eta}} :
\widetilde{U} \to \mathfrak{t}\otimes \tilde{u}^{*}K_{\Sigma}$
corresponding to the canonical section $\widehat{\underline{\Omega}}
\in H^{0}(\widehat{\mathcal{Q}},\Omega^{2}_{\hat{\bq}}\otimes
(\tilde{u}\circ \bq)^{*} K_{\Sigma})$. Using the cut-and-paste
construction of $\widehat{\mathcal{Q}}$ from \cite{balazs-artin}
one can check that the universal period map $\hat{\underline{\eta}}$
is given by the tautological section. Indeed, if we choose a local
frame of $V$ on an open $D \subset \Sigma$ and if we write $\zeta$ for
the corresponding local frame of $K_{\Sigma} \cong \wedge^{2} V$, then
over the local patch $D \subset \Sigma$ we have $\widetilde{\bT}_{|D}
\cong D \times \mathfrak{t}$, $\widehat{\mathcal{Q}}_{|D} \cong
D\times \widehat{Y}$, where $\widehat{Y} \to \mathfrak{t}$ is the
Brieskorn-Grothendieck simultaneous resolution of the versal deformation
family of $\mathbb{C}/\Gamma$.  In these terms we have
$\widehat{\underline{\Omega}} = p_{D}^{*}\zeta \otimes
p_{\widehat{Y}}^{*} \omega$, where $\omega \in
\Omega^{2}_{\widehat{Y}/\mathfrak{t}}$ is the canonical fiberwise
symplectic form on $\widehat{Y}$. Now the statement follows by the
well known fact \cite[Section~4]{sb} that the period map $\mathfrak{t} \to
\mathfrak{t}$ given by $\omega$ is proportional to the identity, and by the
invariance-under-gluing statement of \cite{balazs-artin}. \ \hfill
$\Box$

\newcommand{\etalchar}[1]{$^{#1}$}


\begin{thebibliography}{DDD{\etalchar{+}}05}

\bibitem[AMV02]{amv} M.~Aganagic, M.~Marino and C.~Vafa.  \newblock
All Loop Topological String Amplitudes From Chern-Simons Theory
\newblock {\em Commun. Math. Phys.} 247:467-512 2004.

\bibitem[AKMV03]{akmv} M.~Aganagic, A.~Klemm, M.~Marino and C.~Vafa.
\newblock The Topological Vertex.  \newblock {\em Commun. Math. Phys.}
254:425-478, 2005.

\bibitem[Che55]{chevalley} C.~Chevalley.  \newblock Invariants of
finite groups generated by reflections.  \newblock {\em
Amer. J. Math.}, 77:778--782, 1955.

\bibitem[Del72]{deligne-hodge2} P.~Deligne.  \newblock {Th\'{e}orie}
de {Hodge} {II}.  \newblock {\em Publications Math\'{e}matiques de
l'{I.}{H.}{E.}{S.}}, 40:5--57, 1972.

\bibitem[DFG02]{dfg1} D.-E.~Diaconescu, B.~Florea and A.~Grassi.
\newblock Geometric Transitions and Open String Instantons.  \newblock
{\em Adv. Theor. Math. Phys.} 6:619-642 2003.

\bibitem[DFG03]{dfg2} D.-E.~Diaconescu, B.~Florea and A.~Grassi.
\newblock Geometric
Transitions, del Pezzo Surfaces and Open String Instantons.  \newblock
{\em Adv. Theor. Math. Phys.} 6:643-702 2003.


\bibitem[DFS05]{dfs} D.-E.~Diaconescu, B.~Florea and N.~Saulina.
\newblock A Vertex Formalism for Local Ruled Surfaces.  \newblock
2005, hep-th/0505192.

\bibitem[DDD{\etalchar{+}}05]{dddhp} D.-E. Diaconescu, R.~Dijkgraaf,
R.~Donagi, C.~Hofman, and T.~Pantev.  \newblock Geometric transitions
and integrable systems.  \newblock 2005, hep-th/0506196.

\bibitem[DV02a]{dv:matrix} R. Dijkgraaf and C. Vafa.  \newblock
Matrix Models, Topological Strings, and Supersymmetric Gauge Theories.
\newblock {\em Nucl. Phys.}, B644:3-20, 2002.

\bibitem[DV02b]{dv:geom} R. Dijkgraaf and C. Vafa.  \newblock On
Geometry and Matrix Models.  \newblock {\em Nucl. Phys.} B644:21-39,
2002.

\bibitem[DG02]{ron-dennis} R.~Donagi and D.~Gaitsgory.  \newblock The
gerbe of {H}iggs bundles.  \newblock {\em Transform. Groups},
7(2):109--153, 2002.

\bibitem[DM96]{dm} R.~Donagi and E.~Markman.  \newblock Cubics,
integrable systems, and {C}alabi-{Y}au threefolds.  \newblock In {\em
Proceedings of the Hirzebruch 65 Conference on Algebraic Geometry
(Ramat Gan, 1993)}, volume~9 of {\em Israel Math. Conf. Proc.}, pages
199--221, Ramat Gan, 1996. Bar-Ilan Univ.

\bibitem[Don95]{cameral} R.~Donagi.  \newblock Spectral covers.
\newblock In {\em Current topics in complex algebraic geometry
(Berkeley, CA, 1992/93)}, volume~28 of {\em
Math. Sci. Res. Inst. Publ.}, pages 65--86.  Cambridge Univ. Press,
Cambridge, 1995.

\bibitem[Fal93]{faltings} G.~Faltings.  \newblock Stable {$G$}-bundles
and projective connections.  \newblock {\em J. Algebraic Geom.},
2(3):507--568, 1993.

\bibitem[Ive86]{iversen} B.~Iversen.  
\newblock {\em Cohomology of
sheaves}.  \newblock Universitext. Springer-Verlag, Berlin, 1986.

\bibitem[LLZ03]{llz} C.-C.M.~Liu, K.~Liu and J.~Zhou.  \newblock A
Proof of a Conjecture of Marino-Vafa on Hodge Integrals.  \newblock
{\em J. Diff. Geom.} 65:289-340, 2004; \newblock A Formula of
Two-Partition Hodge Integrals.  \newblock 2003, math.AG/0310272.

\bibitem[LLLZ04]{lllz} J.~Li, C.-C.M.~Liu, K.~Liu and J. Zhou.
\newblock A Mathematical Theory of the Topological Vertex \newblock
2005, math.AG/0408426.

\bibitem[Slo80]{slodowy} P.Slodowy. 
\newblock Simple singularities and simple algebraic groups.
\newblock Lecture Notes in Math. {\bf 815} (1980), Springer Verlag.

\bibitem[Sze04]{balazs-artin} B.~Szendr\"{o}i.
\newblock Artin group actions on derived categories of threefolds.
\newblock  J. Reine Angew. Math.  572  (2004), 139--166.


\bibitem[Sze05]{balazs-fiber} B.~Szendr\"{o}i.
\newblock Sheaves on fibered threefolds and quiver sheaves.
\newblock  preprint, math.AG/0506301.

\bibitem[ShBa01]{sb} N.~Shepherd-Barron.
\newblock Simple groups and simple singularities. 
\newblock Israel J. of Math. {\bf 123} (2001), 179-188.
\end{thebibliography}

\end{document}